# Study of the surface lattice resonance on basis orientation for achieving ultrahigh quality factor > 12000

C. Liu, Joshua T.Y. Tse, and H.C. Ong[1]

Department of Physics, The Chinese University of Hong Kong, Shatin, Hong Kong, People's Republic of China

Periodic nanoparticle arrays can support surface lattice resonances (SLRs), which arise from the hybridization between localized surface plasmons (LSPs) and diffractive Rayleigh anomalies (RAs). In contrast to LSPs, SLRs enjoy a much higher quality (Q) factor. As the Q factor depends on many system parameters, a good understanding of them is essential for optimization. Here, we study the dependence of the Q factor of SLR from 2D Au nanorod arrays on nanorod orientation. It is found the Q factor is 112 when the nanorod is lying perpendicular to the incident plane but gradually increases by more than 13 times to 1460 upon rotating it azimuthally. The increase of the Q factor is due to the interplay between the coupling strength and the frequency detuning between the LSP and RA as well as the decay rate of the LSP. By optimizing these parameters, we can achieve a Q factor reaching over 12000, which is 6 times higher than the best reported so far.

1) Email: hcong@phy.cuhk.edu.hk

Single nanoparticles support localized surface plasmons (LSPs), which produce strong electromagnetic field enhancement by confining the field energy in an extremely small mode volume [1]. Although the field intensity can be increased by one to two orders of magnitude relative to the excitation, LSPs usually suffer from a low quality (Q) factor, typically < 10, due to their Ohmic absorption loss and Rayleigh or Mie radiation damping [2]. To further boost the field strength, the Q factor must be improved to the level comparable to high finesse cavities such as Fabry-Perot resonators and whispering-gallery mode microcavities that can have the Q factors up to 10000 or even more [3,4].

Nanoparticles arranged in a lattice form can support diffractive Rayleigh anomalies (RAs) that hybridize with LSPs to yield the so-called surface lattice resonances (SLRs) [5-8]. SLRs can have a much higher Q factor, reaching over 2000 for the best reported so far [9]. High Q SLRs can find many applications in linear and nonlinear optics such as surface plasmon resonance biosensing [10], surface-enhanced Raman scattering [11], higher order harmonic generations [12], up-conversion [13], lasing [14] and many more. However, engineering the Q factor of SLRs to suit for an application is a complicated process because it involves many parameters that require fine tuning. To date, it has been reported that the geometry and material of the nanoparticle, the number of nanoparticles in the unit cell, and the substrate and superstrate environments all take part in determining the Q factor [6,9,15-20]. Therefore, how one can parametrize the Q factor and then coordinate all the parameters to yield the best result has become a field of interest.

Recently, we have formulated the Q factor of SLR based on temporal coupled mode theory (CMT) and achieved a Q factor over 600 from dimerized nanodisk arrays [21]. However, dimerization is not the only way to achieve ultrahigh Q factor. Here, we study the effects of nanoparticle orientation on the Q factor of monomer arrays by CMT and finite-difference time-domain (FDTD) simulation. By using nanorod as the basis, we azimuthally rotate the nanorod in a 2D square lattice and find the Q factor increases from 112 to 1460 when the major axis of the nanorod is swiveled closer and closer to the incident plane taken along the Γ-X direction. Such an increase is due to the decrease of the interaction strength between the LSP and RA. As a result, by optimizing the interaction strength and the frequency detuning between the LSP and RA, the decay rate of the LSP, as well as the excitation configuration, we have achieved a Q factor over 12,000 with strong light absorption of 0.36.

In Fig. 1(a), the plane-view 2D square lattice unit cell that has period P = 400 nm and rectangular nanorod with length L = 100 nm, width W = 50 nm, and height H = 50 nm is surrounded by refractive index n = 1.5 environment. The major axis of the nanorod is azimuthally rotated, with respect to the y-axis, from $\rho = 0°$ to 90° with a step size of 15°. A linearly polarized light is incident along the x-axis, or the Γ-X direction, as indicated by the green arrow. As an illustration, the 45° polarized specular reflectivity contour mappings of $\rho$ = 0°, 30°, 60°, and 90° arrays taken along the Γ-X direction are calculated as a function of incident angle θ in Fig 1(b) – (e). They show the nondispersive longitudinal LSP (solid line) is located at $\lambda \approx 780$ nm whereas the dispersive (-1,0) RA (dashed line) is calculated by using the phase-matching equation given as [6,21]:

$$(n/\lambda)^2 = (\sin\theta/\lambda + n_x/P)^2 + (n_y/P)^2, \qquad (1)$$

where $(n_x, n_y)$ is the mode order of the RA. From the $\rho = 0°$, 30° and 60° mappings, we see an avoided crossing occurs at $\theta \approx 16°$ due to the hybridization between the LSP and the (-1,0) RA, yielding an energy band gap as well as two upper and lower coupled bands [6,7]. Conventionally, the long wavelength branch of the coupled bands that follows closely with the RA is known as (-1,0) SLR and it exhibits high Q factor [6]. We notice from the mappings that the LSP-RA hybridization is the strongest when $\rho = 0°$ but is null when $\rho = 90°$. The weakening of the coupling strength is due to the mismatch between the radiations of LSP and the (-1,0) RA. In general, the radiations from the nanoparticles should be aligned with the propagation direction of the RA in order to couple all LSPs into the collective SLR [22]. Therefore, for Γ-X propagating RA, the LSP radiations must point in the x-direction. In other words, the major axis of the nanorods should be oriented in the y-direction, or $\rho = 0°$, under the TE-polarized incidence for yielding the strongest coupling. $\rho = 90°$ nanorods excited by TM-polarized light lead to radiations orthogonal to the RA propagation direction, resulting in no coupling. Remarkably, closer examination of the mappings reveals the linewidth of the (-1,0) SLR becomes narrower when $\rho$ increases. We then extract the reflectivity spectra taken at $\theta = 24°$ with the SLR reflection peaks locating at ~ 891 - 899 nm in Fig. 2(a) for comparison. In fact, one sees the linewidth decreases from 8 nm to 0.6 nm when $\rho$ increases from 0° to 75°, leading to the Q factor, defined as $\omega/\Delta\omega$, increases from 112 by 13 times to 1460 as displayed in Fig. 2(b).

To elucidate such dependence, we formulate the interactions between the LSP and two TE- and TM-polarized RAs by CMT [23-25]. The dynamics of three modes are formulated as [21]:

$$\frac{d}{dt}\begin{bmatrix} a_{LSP} \\ a_{TE} \\ a_{TM} \end{bmatrix} = i \begin{bmatrix} \tilde{\omega}_{LSP} & \Omega_{TE} & \Omega_{TM} \\ \Omega_{TE} & \tilde{\omega}_{TE} & 0 \\ \Omega_{TM} & 0 & \tilde{\omega}_{TM} \end{bmatrix} \begin{bmatrix} a_{LSP} \\ a_{TE} \\ a_{TM} \end{bmatrix} + K^T |S\rangle, \quad (2)$$

where $a_{LSP,TE,TM}$ and $\tilde{\omega}_{LSP,TE,TM} = \omega_{LSP,TE,TM} + i\Gamma_{LSP,TE,TM}/2$ are the mode amplitudes and the complex angular frequencies of LSP, TE- and TM-RAs, where $\omega_{LSP,TE,TM}$ and $\Gamma_{LSP,TE,TM}$ are the corresponding angular frequencies and decay rates. The coupling constants between LSP and TE-RA and between LSP and TM-RA are defined as $\Omega_{TE}$ and $\Omega_{TM}$, respectively. Since the TE- and TM-RAs are orthogonal, degenerate, and almost lossless, they can be considered as $\tilde{\omega}_{TE,TM} = \omega_{RA}$. Therefore, the interaction between the modes is mostly near-field in nature and both $\Omega_{TE,TM}$ are real values [21]. $K = \begin{bmatrix} \kappa\cos\rho & \kappa\sin\rho & \kappa\cos\rho & \kappa\sin\rho \\ 0 & 0 & 0 & 0 \\ 0 & 0 & 0 & 0 \end{bmatrix}$ and $|S\rangle = \begin{bmatrix} s_{TE}^R & s_{TM}^R & s_{TE}^T & s_{TM}^T \end{bmatrix}^T$ are the complex in-coupling matrix and the incident power amplitude vector for TE- and TM-polarizations, where $\kappa$ is the in-coupling constant to the LSP that carries a sinusoidal $\rho$ dependence for TE- and TM-polarizations and the R and T superscripts in the power amplitudes stand for the reflection and transmission side incidences. As the RAs are nonradiative, their in-coupling constants are zero. We solve the homogenous part of Eq. (2) to obtain the complex eigenfrequencies, $\tilde{\omega}_{1-3}$, to be:

$$\tilde{\omega}_1 = \omega_{RA} \text{ and } \tilde{\omega}_{2,3} = \omega_{2,3} + i\frac{\Gamma_{2,3}}{2} = \frac{(\tilde{\omega}_{LSP} + \omega_{RA})}{2} \pm \sqrt{\left(\frac{\tilde{\omega}_{LSP} - \omega_{RA}}{2}\right)^2 + \Omega_{TE}^2 + \Omega_{TM}^2}. \quad (3)$$

While $\tilde{\omega}_1$ is RA-like, $\tilde{\omega}_{2,3}$ are the upper and lower SLR bands. In fact, the long wavelength SLR is $\tilde{\omega}_3$ and its Q factor, defined as $\omega_3/\Gamma_3$, can be approximated as [21]:

$$\frac{4}{\Gamma_{LSP}}\left(\omega_{RA}\left(\frac{(\omega_{RA} - \omega_{LSP})^2 + (\Gamma_{LSP}/2)^2}{4(\Omega_{TE}^2 + \Omega_{TM}^2)} + 2\right) - \omega_{LSP}\right). \quad (4)$$

One sees Eq. (4) provides a simple analytical form for engineering the Q factor and it depends on the frequency detuning $\omega_{RA} - \omega_{LSP}$ between the RA and LSP, the decay rate $\Gamma_{LSP}$ of the LSP, and the coupling constants $\Omega_{TE,TM}$. For our case, however, upon varying $\rho$, both $\tilde{\omega}_{LSP}$ and $\omega_{RA}$ are not expected to be changed, and the variation of the Q factor is primarily due to the change of $\Omega_{TE}^2 + \Omega_{TM}^2$.

We will determine $\Omega_{TE,TM}$ accordingly. The spectral positions $\lambda_{2,3}$ and the linewidths $\Delta\lambda_{2,3}$ of two coupled bands from all the arrays are extracted by using Fano function fitting in Fig. 3(a) – (d). We see from $\lambda_{2,3}$ the band gap size decreases with increasing ρ and eventually becomes zero when ρ = 90°. At the same time, $\Delta\lambda_{2,3}$ cross at the band gap, verifying the LSP-RA couplings are near-field in nature [26]. As the gap size reflects the coupling strength, it weakens when ρ increases. The $\lambda_{2,3}$ plots are then fitted with the real part of $\tilde{\omega}_{2,3}$ in Eq. (3) to determine $\Omega_{TE}^2 + \Omega_{TM}^2$ by assuming $\tilde{\omega}_{LSP}$ = 1.59 + i0.12 eV and $\omega_{RA}$ following the phase-matching equation given in Eq. (1). The best fits are shown as the dashed lines in Fig. 3 with the fitted $\Omega_{TE}^2 + \Omega_{TM}^2$ being summarized in Fig. 3(e), which indicates the coupling decreases nonlinearly with increasing ρ.

We then deconvolute $\Omega_{TE}$ and $\Omega_{TM}$ with the aid of the CMT and the simulated field patterns. We solve the eigenvector for $\tilde{\omega}_3$ to be $\left[\omega_{RA} - \tilde{\omega}_3 \quad -\Omega_{TE} \quad -\Omega_{TM}\right]^T$ so that the mode amplitude $a_3$ carries the form of $\beta\left[(\omega_{RA} - \tilde{\omega}_3)a_{LSP} - \Omega_{TE}a_{TE} - \Omega_{TM}a_{TM}\right]$, where $\beta = 1/\sqrt{(\omega_{RA} - \tilde{\omega}_3)^2 + \Omega_{TE}^2 + \Omega_{TM}^2}$ is the normalization constant. The SLR fields thus are the superposition of the fields from the LSP, TE- and TM-RAs. Given the fields of the LSP are spatially confined around the nanorod in the x-y-plane, examining the propagating fields far away from the nanorod in the y- and the x- and z-directions may provide the information of $\Omega_{TE}$ and $\Omega_{TM}$. The near-field patterns of the (-1,0) SLRs in the x-, y- and z-directions for the arrays taken at θ = 24° under 45° polarized light are simulated in Fig. 4(a) – (l). Apparently, for ρ = 0°, 30°, and 60° arrays, other than the strong localized fields around the nanorod in the x- and y-directions, their y-components are strong at x = ±200 nm of the unit cell, but both the x- and z-components are negligible, indicating the SLR fields are predominately TE-like without any superposition of $a_{TM}$, suggesting $\Omega_{TM} \approx 0$. On the other hand, for the ρ = 90° array, the fields display only the localized field characters around the nanorod and are weak at x = ±200 nm for all components. It shows only the LSP is present without interacting with the RAs $a_{TE}$ and $a_{TM}$, and thus both $\Omega_{TE,TM} = 0$. Therefore, it is reasonable to argue, for the nanorod lying on the surface, $\Omega_{TM} = 0$ all the time regardless of ρ, and the dependence shown in Fig. 3(e) is solely due to $\Omega_{TE}$. We also find $\Omega_{TE}$ follows a $\cos\rho$ dependence as indicated by the dashed line, indicating $\Omega_{TE}$ scales with the y-component of the dipole moment induced by the nanorod. Fig. 2(c) shows the plots of the Q factor against $1/\Omega_{TE}^2$, clearly demonstrating a linear behavior as in Eq. (2).

Once the dependence of the Q factor on the coupling constants has been studied, we are in the position of rationally designing nanoparticle arrays for realizing ultrahigh Q SLR with strong light absorption. In prior to presenting the simulation results, we first discuss the effects of the frequency detuning $\omega_{RA} - \omega_{LSP}$ and the decay rate of LSP $\Gamma_{LSP}$ in Eq. (4) for enhancing the Q factor. We see the detuning is a trivial parameter which does not require much engineering [6,7]. While the LSP is a localized mode in which the $\omega_{LSP}$ does not depend on θ, the RA is dispersive with its $\omega_{RA}$ strongly dependent on θ following the phase matching equation given in Eq. (1). Therefore, one simply needs to increase the incident angle to enlarge the detuning. On the other hand, the decay rate $\Gamma_{LSP}$ deserves some attention. Sönnichsen et al have carried out a systematic study on the Q factors of single nanospheres and nanorods and find nanorods exhibits 10 times higher Q factor than nanospheres due to the suppression of interband absorption damping [27]. In addition, nanorods with higher aspect ratio have higher Q factor. Other than the shape, the size of the nanoparticles is also important because it governs the Rayleigh or Mie radiation damping [3]. As a result, using small size and high aspect ratio nanorods as the building block can effectively reduce both the absorption and radiative decay rates at the same time.

The excitation configuration is of importance to facilitate strong light absorption and thus the field energy. In fact, the CMT provides the conditions for maximizing the energy of the SLR. Assuming the SLRs are excited from the reflection side with $|S\rangle = \begin{bmatrix} \cos\gamma & e^{i\delta}\sin\gamma & 0 & 0 \end{bmatrix}^T$, where γ and δ are the polarization angle and the phase difference between the TE- and TM-incidences, after the diagonalization of Eq. (2) [21,28], we find $\frac{da_3}{dt} = i\tilde{\omega}_3 a_3 - \frac{\kappa\Omega_{TE}}{\sqrt{(\tilde{\omega}_{LSP} - \omega_{RA})^2 + 4\Omega_{TE}^2}}\left(\cos\gamma\cos\rho + e^{i\delta}\sin\gamma\sin\rho\right)$ and the SLR mode energy $|a_3|^2$ at resonance, i.e. ω = ω₃, is:

$$\frac{(2\kappa\Omega_{TE})^2}{\left|\sqrt{(\tilde{\omega}_{LSP} - \omega_{RA})^2 + 4\Omega_{TE}^2}\right|^2} \frac{\left|\cos\gamma\cos\rho + e^{i\delta}\sin\gamma\sin\rho\right|^2}{\Gamma_3^2}. \tag{5}$$

As a result, one can tune the incidence γ and δ for optimizing $|a_3|^2$, or the absorption.

We then carry out FDTD simulations on two series of arrays to verify our proposition. The first series has 2D Au nanorod arrays with $P_x$ = 500 nm and $P_y$ = 100 in the x- and y-directions and L = 60 nm, W = 20 nm, and H = 20 nm. The aspect ratio is 3 to increase the Q factor of a single nanorod [27]. At the same time, we also reduce $P_y$ to 100 nm to increase the nanorod density and enhance the extinction ratio [9]. Fig. 5(a) shows the reflectivity spectra taken under 45º linearly polarized light at θ = 22º along the Γ-X direction for different ρ. The linewidth of the (-1,0) SLR peak at 1060.4 – 1063.6 nm is found to decrease gradually with increasing ρ, yielding the highest Q factor to be 16177 at ρ = 75º, as summarized in Fig. 5(b). This Q factor is at least 7 times higher than the best reported to date and is expected to increase further at longer wavelength when $\omega_{RA} - \omega_{LSP}$ increases. On the other hand, for the second series, we simulate $P_x$ = 500 nm and $P_y$ = 80 nm and L = 60 nm, W = 20 nm, and H = 20 nm with ρ = 70º under different incident conditions. Fig. 5(c) contour plots the resonant absorption peak intensity at 1060.53 nm taken under different γ and δ. It shows the strongest absorption of 0.36 occurs at γ = 46 deg and δ = 0.5 deg with the Q factor = 11516. As the system supports both reflection and transmission, we expect the absorption will become stronger when under two-port excitation in which light is incident from both the reflection and transmission sides [29,30].

In summary, we have studied the dependence of the Q factor of SLRs supported by 2D Au nanorod arrays on nanorod orientation. It is found the Q factor is the lowest when the nanorod is perpendicular to the incident plane along the Γ-X direction but increases when the nanorod is swiveled away. The increment is due to the weakening of the interaction between the LSP and the RA, which inversely scales with the Q factor, as formulated by temporal CMT. More importantly, based on the CMT, we have systemically optimized the coupling strength and the frequency detuning between the LSP and the RAs, the decay rate of the LSP, and the excitation configuration to achieve a Q factor as high as 11516 with the absorption = 0.36.

This research was supported by the Chinese University of Hong Kong through Area of Excellence (AoE/P-02/12) and Innovative Technology Fund Guangdong-Hong Kong Technology Cooperation Funding Scheme (GHP/077/20GD) and Partnership Research Program (PRP/048/22FX).

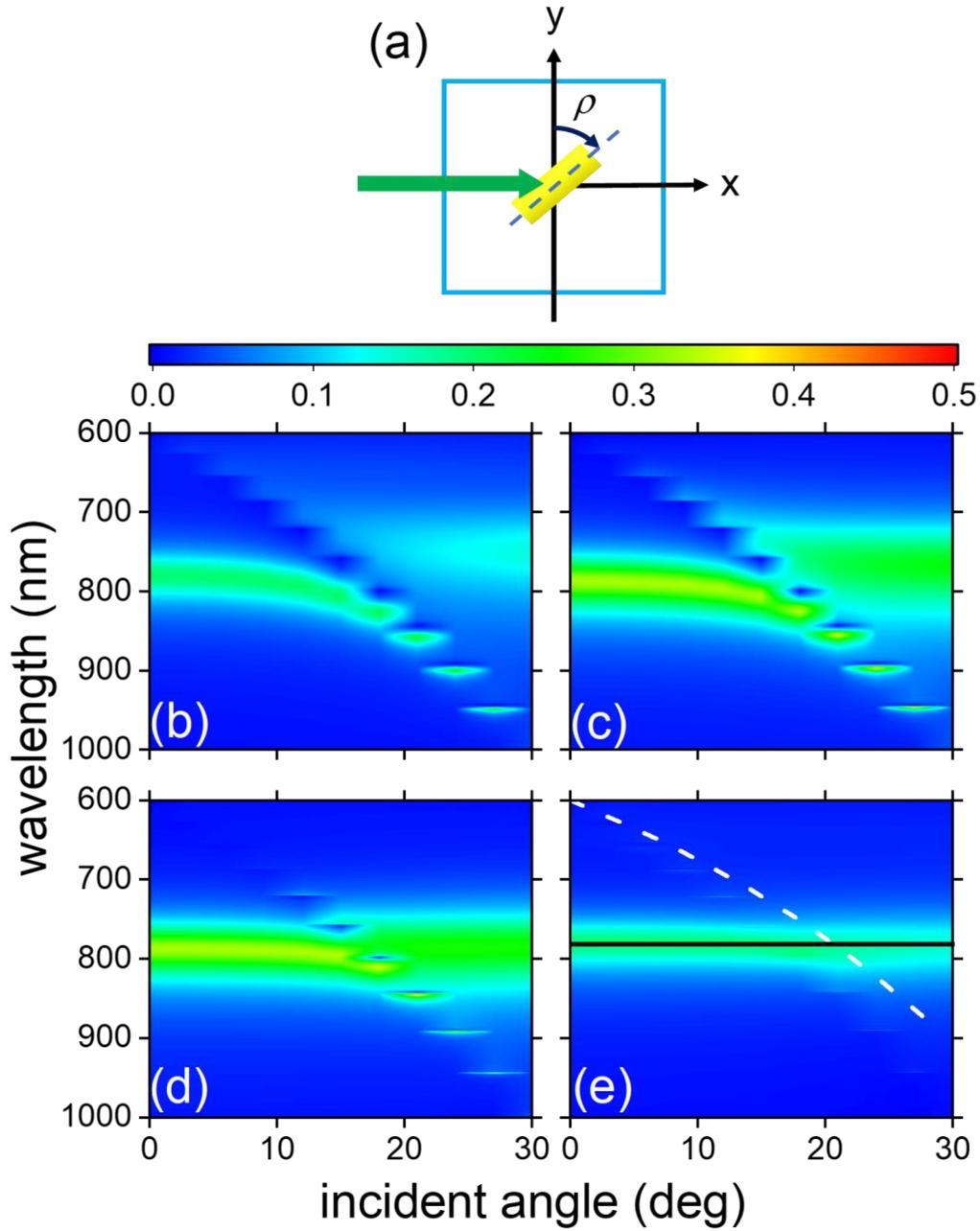

Fig. 1. (a) The FDTD unit cell with the nanorod rotated azimuthally with respect to the y-axis by ρ and is excited by a polarized light (green arrow) along the Γ-X direction. The θ-resolved reflectivity mappings taken along the Γ-X direction for ρ = (a) 0°, (b) 30°, (c) 60° and (d) 90° excited by a 45° linearly polarized light. The dashed line is the (-1,0) RAs and the solid line is the LSP. No coupling occurs when ρ = 90°.

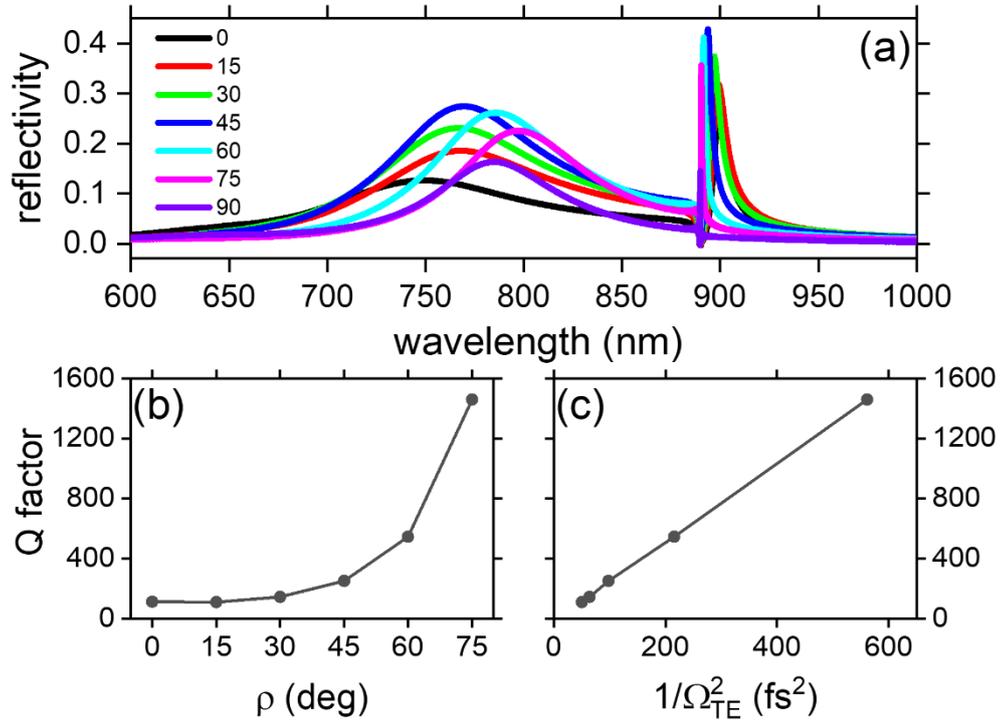

Fig. 2. (a) The reflectivity spectra of the square lattice arrays taken at θ = 24° for different ρ. The (-1,0) SLR reflection peaks are visible at 891 – 899 nm and become narrower when ρ increases. (b) The plot of the corresponding SLR Q factor against ρ. (c) The plot of the Q factor against $1/\Omega_{TE}^2$, showing a linear behavior.

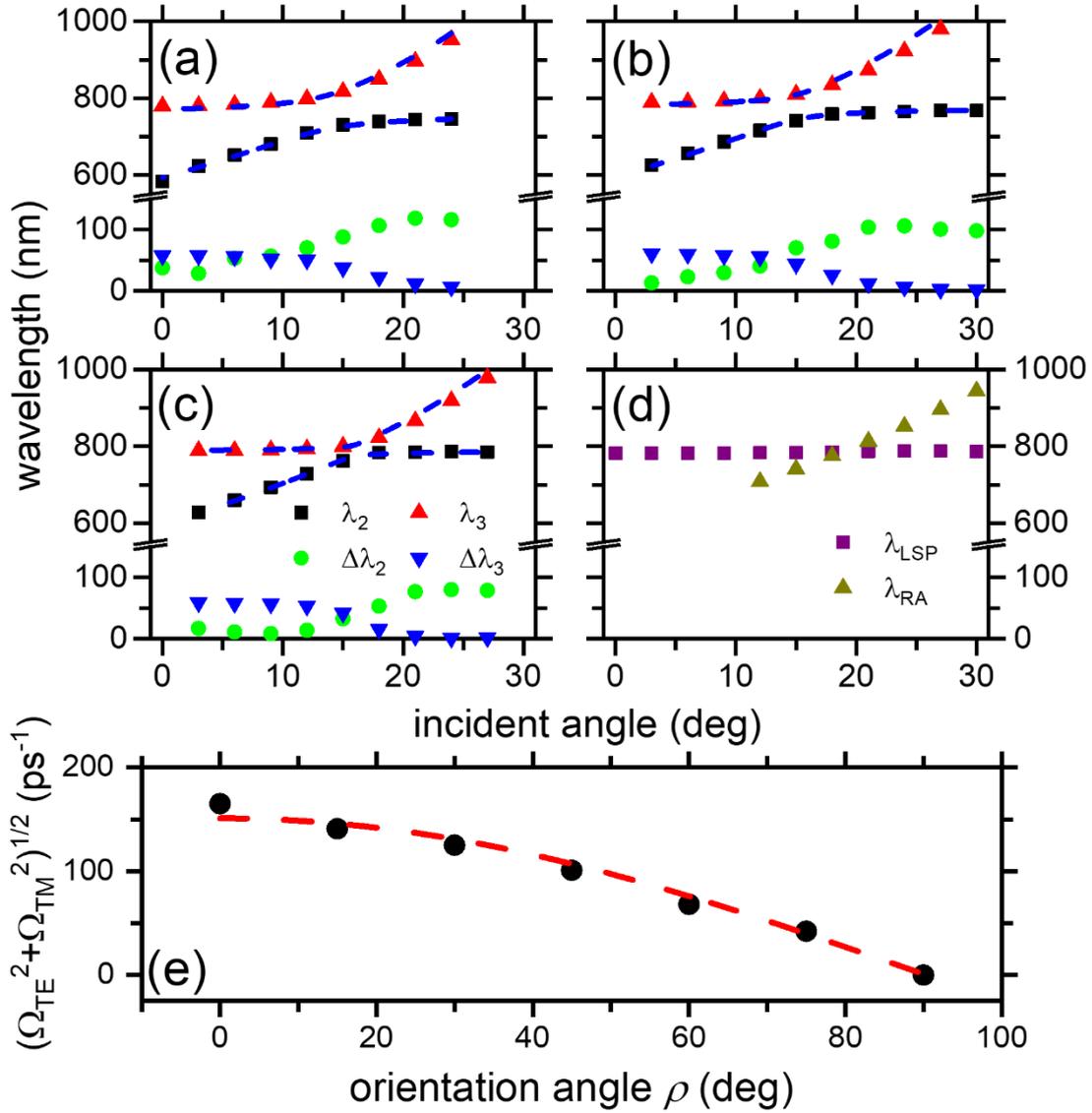

Fig. 3. The plots of the spectral positions $\lambda_{2,3}$ and the linewidths $\Delta\lambda_{2,3}$ of the upper and lower bands for $\rho$ = (a) 0°, (b) 30°, and (c) 60° as a function of θ. The dashed lines are the best fits determined by CMT. (d) The plots of the spectral positions $\lambda_{LSP,RA}$ of the LSP and RA for $\rho$ = 90° array as a function of θ. (e) The plots of $\sqrt{\Omega_{12}^2+\Omega_{13}^2}$ against ρ and the dashed line is the cosρ best fit.

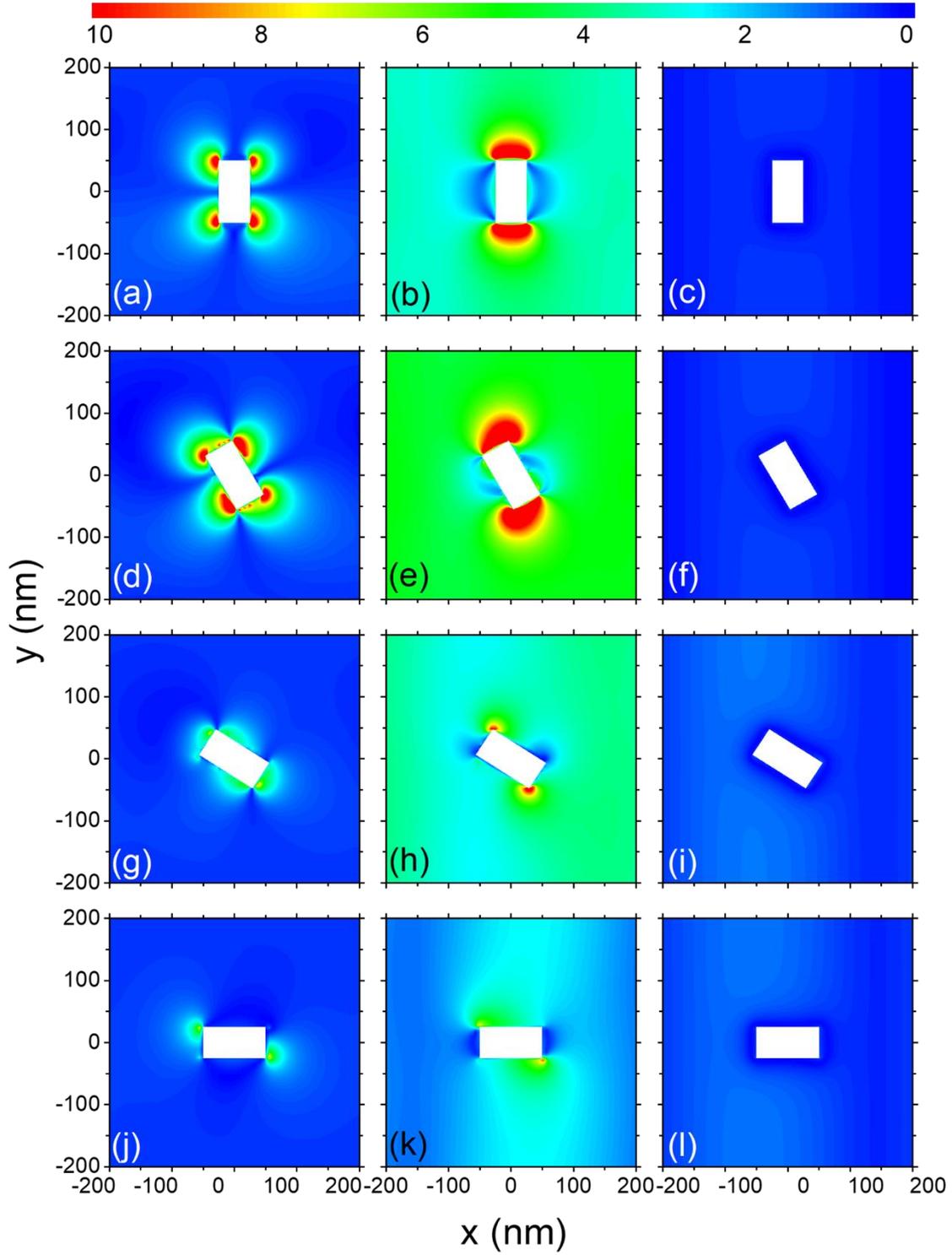

Fig. 4. The near-field patterns of (-1,0) SLR with (a) – (c) $\rho = 0°$, (d) – (f) $\rho = 30°$, (g) – (i) $\rho = 60°$, and (j) – (l) $\rho = 90°$. The first, second and third columns are the x-, y-, and z-components of the field intensities. The white blank areas are the nanorods.

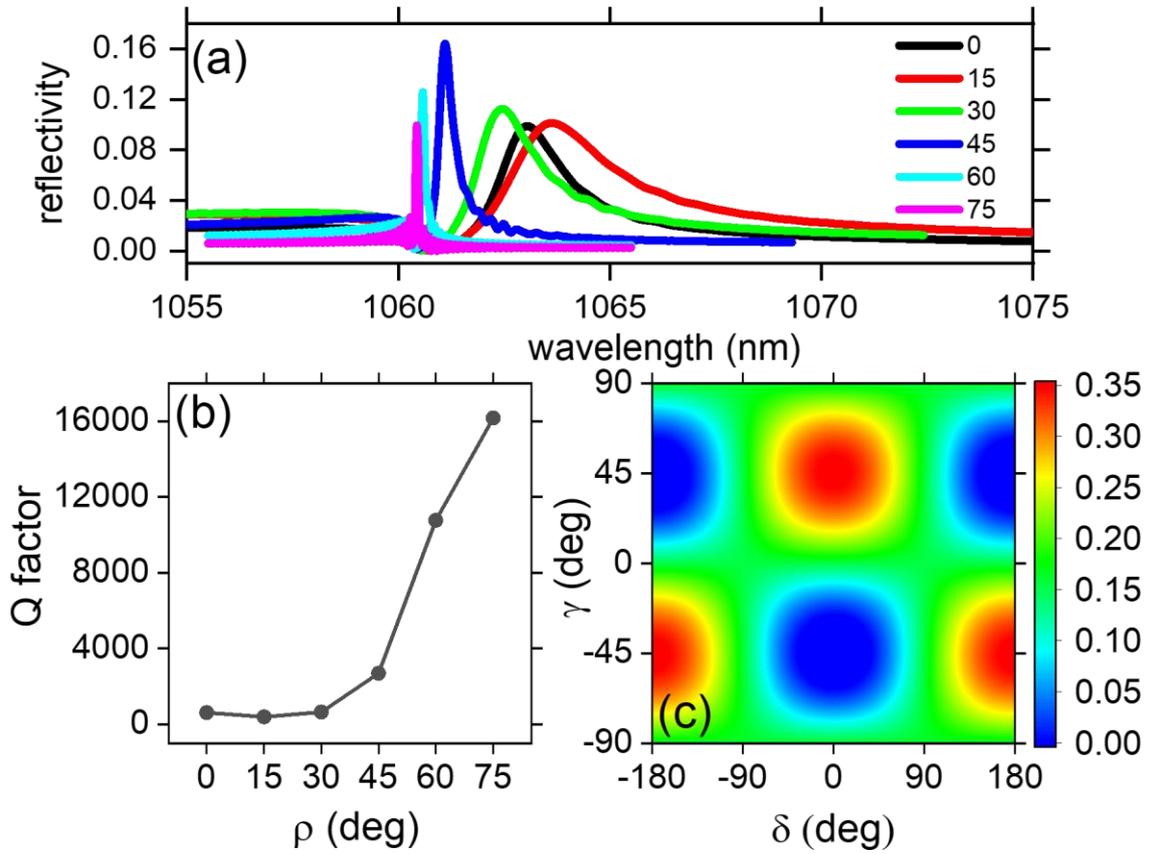

Fig. 5 (a) The reflectivity spectra of the first series rectangular lattice arrays taken at θ = 22° for different ρ. The (-1,0) SLR reflection peaks are at 1060.4 – 1063.6 nm. (b) The plot of the corresponding SLR Q factor against ρ. (c) The resonant absorption peak intensity mapping of the second rectangular lattice array at 1060.6 nm taken under different γ and δ.